\documentclass[11pt,a4paper]{article}
\usepackage[margin=1in]{geometry}
\usepackage[T1]{fontenc}
\usepackage[utf8]{inputenc}
\usepackage{lmodern}
\usepackage{microtype}
\usepackage{amsmath,amssymb,amsthm}
\usepackage{graphicx}
\usepackage{array,tabularx,booktabs}
\usepackage{xcolor}
\usepackage{float}
\usepackage{pgfplots}
\pgfplotsset{compat=1.16}
\usepackage[hidelinks]{hyperref}
\usepackage{xurl}
\usepackage{enumitem}
\usepackage{setspace}
\definecolor{reviewA}{RGB}{0,0,0}
\definecolor{reviewB}{RGB}{0,0,0}
\definecolor{reviewC}{RGB}{0,0,0}
\newcommand{\N}{\mathcal{N}}
\newcommand{\W}{\mathcal{W}}
\newcommand{\T}{\mathcal{T}}
\newcommand{\Tp}{\mathcal{T}^{+}}
\newcommand{\D}{\mathcal{D}}
\newcommand{\Sset}{\mathcal{S}}
\newcommand{\Gset}{\mathcal{G}}
\newcommand{\R}{\mathcal{R}}
\newcommand{\Lset}{\mathcal{L}}
\newcommand{\TN}{\mathcal{T}^{N}}
\newcommand{\Pset}{\mathcal{P}}
\newcommand{\eps}{\varepsilon}
\newcommand{\APAReference}[1]{\par\noindent\hangindent=1.5em\hangafter=1 #1\par\smallskip}
\newenvironment{simpleReferences}{\section*{References}\small\sloppy\setlength{\emergencystretch}{3em}\setlength{\parindent}{0pt}}{}
\title{Bi-Objective Receding-Horizon Scheduling of Nurses and Hospital Service Robots}
\author{Khatere Ghorbani-Moghadam$^{1}$ \and Roya Khalili-Amirabadi$^{2,*}$}
\date{}
\begin{document}
\maketitle
\begin{center}
\small $^{1}$Mosaheb Institute of Mathematics, Kharazmi University, Tehran, Iran\\
\href{mailto:k.ghorbani@khu.ac.ir}{k.ghorbani@khu.ac.ir}\\[4pt]
$^{2}$Department of Applied Mathematics, Ferdowsi University of Mashhad, Mashhad, Iran\\
\href{mailto:roya.khalili.a@gmail.com}{roya.khalili.a@gmail.com}\\[3pt]
$^{*}$Corresponding author
\end{center}
\begin{abstract}
This paper studies the joint scheduling of nurses and hospital service robots as a bi-objective mixed-integer linear programming problem. The formulation represents heterogeneous nursing skills, alternative skill-coverage modes, nurse availability and temporal rules, workload balance, robot task capability, charging and battery dynamics, and human supervision. Robot workload relief is credited only when an eligible service is actually executed, and explicit supervision-load variables deduct the corresponding nurse capacity from clinical coverage. The receding-horizon formulation also includes a cross-window night--morning constraint and a terminal battery reserve. The computational study is presented as numerical verification on synthetic hospital-like instances rather than clinical validation. Under the proposed formulation, the eight-nurse reference case gives $Z_1=0.78$ and $Z_2=3.164$. A common-budget comparison on B20 seed~1 yields total shortage $53.70$ for the integrated model, compared with $58.03$ for sequential nurse--robot scheduling and $59.46$ for the nurse-only baseline. Additional experiments examine multiple epsilon budgets on a larger B30 instance, sensitivity to robot-relief/supervision parameters and objective weights, and one-day versus two-day rolling horizons against a static full-horizon reference. The battery and routing representations remain aggregate planning approximations, and the scalability evidence is interpreted accordingly. No patient records or real hospital staff data are used, so the results establish model behavior and comparative numerical evidence rather than practical clinical effectiveness.
\end{abstract}
\medskip

Nurse scheduling is not simply a question of meeting headcount. A workable roster must place appropriately qualified nurses in the right wards and shifts while respecting availability, labor rules, workload balance, preferences, and temporal restrictions. These requirements have been studied extensively in the nurse rostering and personnel scheduling literature (Cheang et al., 2003; Burke et al., 2004; Ernst et al., 2004; Van den Bergh et al., 2013; De Bruecker et al., 2015). In hospital settings, skill mix is especially important because two rosters with the same number of nurses can differ substantially in clinical adequacy when critical-care or senior competencies are scarce.

Service robots introduce a second resource class whose usefulness depends on both task capability and internal state. Hospital robots may assist with transport, specimen delivery, supply movement, disinfection, telepresence, and other non-specialist activities. Reviews of healthcare robotics describe a growing range of such applications (Morgan et al., 2022; Huang et al., 2023), and systems such as Lio demonstrate the practical feasibility of supportive robotic assistance in human-centered care environments (Miseikis et al., 2020). Robotics has also been considered for reducing avoidable exposure and supporting logistics during infectious-disease events (Yang et al., 2020). These systems are not substitutes for specialized nursing care. They have limited capabilities, consume energy, require charging, and may depend on human supervision. Work on mixed human--robot teams also shows that coordination, authority, and workload affect team performance (Gombolay et al., 2015), while autonomous mobile robot planning highlights the close connection between task allocation, energy, and operational control (Fragapane et al., 2021).

The scheduling problem considered here therefore treats nurses and robots within the same ward--shift environment. The model combines multi-level nursing skills and alternative acceptable coverage patterns with separate objectives for service adequacy and operational quality. Robot presence is distinguished from actual task execution, and battery dynamics connect decisions across time. A receding-horizon control layer then allows the schedule to be updated when demand, availability, or robot state changes. The resulting formulation is intended as a decision-support model: robots can relieve selected support workload, but they cannot replace sensitive clinical competencies.

Model predictive control (MPC) provides a natural structure for this type of repeated decision making because it optimizes over a finite horizon, implements the first decision, observes the updated system, and then solves again (Camacho and Bordons, 2007; Borrelli et al., 2017; Qin and Badgwell, 2003). Related work has used mixed-integer predictive control in fleet systems with charging constraints (Zhang et al., 2016). More recent research has also examined learning-assisted predictive control. In particular, RL--MPC architectures have been surveyed as a way to combine explicit constraint handling with data-driven adaptation (Jalaeian Farimani et al., 2026), while experience-replay methods provide a broad set of mechanisms for reusing past interaction data in reinforcement learning (Khalili Amirabadi et al., 2026). Reinforcement learning is not used in the present study; these directions are relevant as possible extensions when reliable operational data become available.

{\color{black}The methodological contribution is an integration contribution rather than a claim of a new generic optimization algorithm. Classical nurse-rostering research focuses on staffing, skills, contractual rules, and preferences (Cheang et al., 2003; Burke et al., 2004; Ernst et al., 2004; Van den Bergh et al., 2013; De Bruecker et al., 2015); healthcare-robot studies mainly characterize applications and human--robot interaction (Morgan et al., 2022; Huang et al., 2023; Miseikis et al., 2020; Gombolay et al., 2015); autonomous mobile-robot work emphasizes task allocation, routing, and energy (Fragapane et al., 2021); and predictive-control studies provide the rolling optimization mechanism (Camacho and Bordons, 2007; Borrelli et al., 2017; Zhang et al., 2016). The present model couples nurse assignments, skill coverage, robot task execution, charging, battery state, and supervision in one ward--shift MILP, while keeping clinical/service shortage separate from operational burden. Accordingly, the contribution lies in the joint formulation and state-coupled scheduling architecture rather than in a new generic MILP solution paradigm.}
Table \ref{tab:litposition} summarizes the positioning of the proposed formulation relative to representative nurse-rostering, healthcare-robotics, mixed human-robot planning, and predictive-control literature streams.
{\color{black}
\begin{table*}[!t]
\centering
\caption{\textcolor{black}{Positioning of the present formulation relative to selected literature streams.}}
\label{tab:litposition}
\small
\renewcommand{\arraystretch}{1.25}% More breathing room in Table 1 only
\setlength{\extrarowheight}{2pt}
{\color{black}
\begin{tabularx}{0.96\textwidth}{>{\raggedright\arraybackslash}p{0.24\textwidth}>{\raggedright\arraybackslash}p{0.30\textwidth}X}
\toprule
Literature stream & Principal elements emphasized & Relation to the present study \\
\midrule
Nurse rostering and skill-based workforce planning (Cheang et al., 2003; Burke et al., 2004; Van den Bergh et al., 2013; De Bruecker et al., 2015) & Staffing coverage, skills, contractual rules, preferences, fairness & Provides the nurse-rostering foundation; the cited representative works do not provide the joint nurse--robot state model studied here. \\
Healthcare/service robotics (Morgan et al., 2022; Huang et al., 2023; Miseikis et al., 2020) & Robot applications, assistance tasks, deployment contexts, human interaction & Establishes feasible hospital robot roles; the cited representative works are not joint nurse-roster optimization models. \\
Mixed human--robot teams and AMR planning (Gombolay et al., 2015; Fragapane et al., 2021) & Human--robot coordination, task allocation, routing, energy, operational control & Motivates coordination and energy constraints; the present model instead uses an aggregate ward--shift representation coupled to clinical staffing. \\
Predictive control/fleet MPC (Camacho and Bordons, 2007; Borrelli et al., 2017; Zhang et al., 2016) & Finite-horizon optimization, state propagation, charging-aware fleet decisions & Provides the rolling optimization mechanism, but not heterogeneous nurse-skill coverage. \\
Present study & Nurse skills, robot service execution, battery/charging, supervision workload, and rolling re-optimization & Integrates these elements in one aggregate ward--shift MILP; the contribution is the joint formulation rather than a new generic MILP algorithm. \\
\bottomrule
\end{tabularx}}
\end{table*}}

{\color{black}The mathematical formulation is numerically verified on a small reference case and then examined on replicated synthetic instances and a seven-day rolling-horizon experiment. The term numerical verification is used rather than practical validation because no real hospital dataset, clinical staffing policy, or robot telemetry dataset is used in this study.}

\section*{2. PROBLEM DESCRIPTION AND NOTATION}

Consider a hospital with nurses $\N$, wards $\W$, planning periods $\T$, days $\D$, clinical and support skills $\Sset$, acceptable skill-coverage modes $\Gset$, service robots $\R$, and robot service types $\Lset$. Let $\T_d\subseteq\T$ be the periods belonging to day $d$, let $\TN\subseteq\T$ denote night periods, and let $\Pset$ contain incompatible ordered shift pairs, such as a night shift followed immediately by a morning shift. Battery states are defined on $\Tp=\{1,\ldots,|\T|+1\}$ so that the terminal state after the final scheduling period is represented explicitly.

A nurse may be assigned to at most one ward in a period and, under the baseline policy used here, to at most one shift per day. Each nurse has an availability profile, a regular-hour threshold, a maximum number of night duties, and a set of skill qualifications. For each ward, period, and skill, the hospital may specify more than one acceptable coverage pattern. For example, a requirement may be met by a smaller number of experts or by a larger number of experienced nurses. One of these modes is selected by the optimization model.

{\color{black}Robots are modeled as supportive resources. A robot can work in at most one ward in a period, charge, or remain idle. When active, it may execute only tasks for which it is qualified, subject to a period task capacity. Its battery is reduced by work and replenished by charging. A robot contributes to nursing workload relief only through an explicitly executed service variable; presence in a ward by itself creates no nursing credit. For specialized ICU, senior, or otherwise sensitive clinical skills, the relief coefficient is set to zero. Human supervision is represented in two ways: a headcount ratio limits simultaneous robot activity, and a continuous supervision-load variable reserves a fraction of assigned nurse capacity so that a supervising nurse is not simultaneously credited as fully available for clinical coverage.}

Table~\ref{tab:sets} summarizes the main sets, and Table~\ref{tab:params} lists the principal parameters.

\begin{table*}[!t]
\centering
\caption{Sets and indices.}
\label{tab:sets}
\begin{tabularx}{0.92\textwidth}{>{\raggedright\arraybackslash}p{0.18\textwidth}X}
\toprule
Symbol & Description \\
\midrule
$\N$ & Nurses, indexed by $n$ \\
$\W$ & Hospital wards, indexed by $w$ \\
$\T$ & Scheduling periods, indexed by $t$ \\
$\Tp$ & Battery-state periods, $\{1,\ldots,|\T|+1\}$ \\
$\D$ & Days; $\T_d\subseteq\T$ contains the periods of day $d$ \\
$\Sset$ & Nursing skill and support categories, indexed by $s$ \\
$\Gset$ & Acceptable skill-coverage modes, indexed by $g$ \\
$\R$ & Hospital service robots, indexed by $r$ \\
$\Lset$ & Robot service types, indexed by $l$ \\
$\TN$ & Night periods \\
$\Pset$ & Incompatible ordered shift pairs $(t,\tau)$ \\
\bottomrule
\end{tabularx}
\end{table*}

\begin{table*}[!t]
\centering
\caption{Main model parameters.}
\label{tab:params}
\small
\begin{tabularx}{0.96\textwidth}{>{\raggedright\arraybackslash}p{0.23\textwidth}X}
\toprule
Parameter & Meaning \\
\midrule
$Q_{nsg}$ & 1 if nurse $n$ is sufficiently qualified for skill $s$ under coverage mode $g$, 0 otherwise \\
$H^g_{wts}$ & Required effective coverage for ward $w$, period $t$, skill $s$, under mode $g$ \\
$M_{wts}$ & Local deactivation constant for the selected-mode constraints, chosen from the corresponding coverage requirements \\
$A_{nt}$ & Nurse availability indicator \\
$h_t$ & Duration of period $t$ in hours \\
$C_n$ & Regular-hour threshold for nurse $n$ \\
$R_n^0$ & Historical work hours carried into the current optimization window \\
$N_n^0$ & Historical night-duty count carried into the current optimization window \\
$B_n^N$ & Maximum allowed night-duty count for nurse $n$ over the relevant horizon \\
$L^{\min}_{wt},L^{\max}_{wt}$ & Minimum and maximum nurse headcount in ward $w$ and period $t$ \\
$Q^E_{ns}$ & 1 if nurse $n$ satisfies the expert threshold for sensitive skill $s$ \\
$E_{wts}$ & Minimum expert presence for a sensitive skill, when required \\
$q_{nwt}$ & Preference or undesirability penalty for assigning nurse $n$ to ward $w$ in period $t$ \\
$A^R_{rl}$ & 1 if robot $r$ can perform service $l$ \\
$K_{rt}$ & Maximum number of robot services executable by robot $r$ in period $t$ \\
$D^R_{wtl}$ & Demand for robot service $l$ in ward $w$ and period $t$ \\
\textcolor{black}{$\alpha_{rls}$} & \textcolor{black}{Nurse-equivalent workload relief generated when robot $r$ actually executes service $l$ for eligible support skill $s$; zero for non-relieving or sensitive clinical skills} \\
\textcolor{black}{$\bar\alpha_{rl}$} & \textcolor{black}{Maximum total nurse-equivalent workload that one execution of service $l$ by robot $r$ may displace across all support-skill categories} \\
$e_{rw}$ & Battery energy used when robot $r$ works in ward $w$ for one period \\
$ch_r$ & Energy added to robot $r$ by one charging period \\
$B_r^{\min},B_r^{\max}$ & Minimum starting level for work and maximum battery capacity \\
$U_r$ & Maximum active work periods per day for robot $r$ \\
$m_r^C$ & Minimum charging periods per day, when imposed \\
\textcolor{black}{$\rho_w^R$} & \textcolor{black}{Maximum number of active robots per assigned nurse in ward $w$ (hard supervision headcount limit)} \\
\textcolor{black}{$d^{sup}_{rw}$} & \textcolor{black}{Fraction of one nurse-period required to supervise active robot $r$ in ward $w$} \\
\textcolor{black}{$\bar\phi_n$} & \textcolor{black}{Maximum fraction of nurse $n$'s assigned period that may be allocated to robot supervision} \\
\textcolor{black}{$B_r^{term}$} & \textcolor{black}{Minimum terminal battery reserve in a receding-horizon window} \\
$\beta_w,\rho_t,\sigma_s,\tau_l$ & Priority weights for ward, period, human skill, and robot service \\
$\eta_R$ & Relative weight of robot-service shortage in $Z_1$ \\
$\lambda_B,\lambda_O,\lambda_N,\lambda_P,\lambda_E,\lambda_C$ & Weights in the operational objective $Z_2$ \\
\bottomrule
\end{tabularx}
\end{table*}

\section*{3. MATHEMATICAL FORMULATION}
This section translates the joint nurse--robot scheduling problem into a
bi-objective mixed-integer linear program. It first defines the decision
variables, then develops the staffing and skill-coverage constraints, the
robot task, charging, battery, and supervision constraints, and finally the
two objective functions used to distinguish service adequacy from operational
burden.

\subsection*{3.1 Decision variables}

The principal binary variables are
\begin{align*}
x_{nwt}&=1 &&\text{if nurse $n$ works in ward $w$ at period $t$},\\
z_{rwt}&=1 &&\text{if robot $r$ is active in ward $w$ at period $t$},\\
v_{rwtl}&=1 &&\text{if robot $r$ executes service $l$ in ward $w$ at period $t$},\\
c_{rt}&=1 &&\text{if robot $r$ charges at period $t$},\\
m_{wtsg}&=1 &&\text{if coverage mode $g$ is selected for $(w,t,s)$}.
\end{align*}
Continuous variables $h^S_{wts}\ge0$ and $u^R_{wtl}\ge0$ represent human-skill and robot-service shortage, respectively. $B_{rt}$ is the battery state of robot $r$ at the start of state period $t\in\Tp$. $R_n$ and $O_n$ denote nurse work hours and overtime, and $R^{\max}$ and $R^{\min}$ denote the maximum and minimum nurse workloads. {\color{black}The variable $\phi_{nwt}\in[0,1]$ denotes the fraction of nurse $n$'s capacity in ward $w$ and period $t$ reserved for robot supervision.}

\subsection*{3.2 Nurse assignment and skill coverage}

A nurse may work in at most one ward in a period and only when available:
\begin{equation}
\sum_{w\in\W}x_{nwt}\le A_{nt},\qquad \forall n,t.
\label{eq:oneward}
\end{equation}
The baseline daily policy permits at most one shift per nurse:
\begin{equation}
\sum_{t\in\T_d}\sum_{w\in\W}x_{nwt}\le1,
\qquad \forall n,d.
\label{eq:oneday}
\end{equation}
Ward headcount is bounded by
\begin{equation}
L^{\min}_{wt}\le\sum_{n\in\N}x_{nwt}\le L^{\max}_{wt},
\qquad \forall w,t.
\label{eq:minmax}
\end{equation}
For every incompatible pair $(t,\tau)\in\Pset$,
\begin{equation}
\sum_{w\in\W}x_{nwt}+\sum_{w\in\W}x_{nw\tau}\le1,
\qquad \forall n,(t,\tau)\in\Pset.
\label{eq:rest}
\end{equation}
Night duties satisfy
\begin{equation}
N_n^0+\sum_{w\in\W}\sum_{t\in\TN}x_{nwt}\le B_n^N,
\qquad \forall n.
\label{eq:night}
\end{equation}

For each ward--period--skill combination, exactly one acceptable coverage mode is selected:
\begin{equation}
\sum_{g\in\Gset}m_{wtsg}=1,
\qquad \forall w,t,s.
\label{eq:modeone}
\end{equation}
{\color{black}Coverage under each candidate mode is enforced by
\begin{align}
&\sum_{n\in\N}Q_{nsg}\bigl(x_{nwt}-\phi_{nwt}\bigr)
+\sum_{r\in\R}\sum_{l\in\Lset}\alpha_{rls}v_{rwtl}
+h^S_{wts} \nonumber\\
&\hspace{20mm}\ge H^g_{wts}-M_{wts}(1-m_{wtsg}),
\qquad \forall w,t,s,g.
\label{eq:skill}
\end{align}
When mode $g$ is selected, Eq.~\eqref{eq:skill} reduces to the corresponding coverage requirement. For nonselected modes, the local big-$M$ term relaxes the constraint. A local value satisfying $M_{wts}\ge\max_{g\in\Gset}H^g_{wts}$ is sufficient and avoids the numerical weakness of an unnecessarily large global constant. The second term is now based on actual task execution $v_{rwtl}$, not robot presence $z_{rwt}$. Consequently, robot presence alone cannot generate nurse-equivalent workload relief, and every credited relief event consumes one unit of the same executed-service capacity that appears in Eq.~\eqref{eq:robotasks}. To prevent a single executed service from being credited independently against several support-skill requirements, the relief coefficients are calibrated to satisfy
\begin{equation}
\sum_{s\in\Sset}\alpha_{rls}\le \bar\alpha_{rl},\qquad \forall r,l,
\label{eq:reliefcap}
\end{equation}
where $\bar\alpha_{rl}$ is the maximum total nurse-equivalent workload that one execution of service $l$ can displace. For sensitive clinical skills, or for robot services that do not displace the corresponding nursing workload, $\alpha_{rls}=0$. This calibration restriction is important: the robot-service demand $D^R_{wtl}$ is interpreted as demand for the same completed support activity whose execution may displace nurse workload, rather than as an additional independent activity. The supervision fraction $\phi_{nwt}$ reduces the effective clinical contribution of an assigned nurse.}

If a hard expert-presence rule is required, it is imposed separately:
\begin{equation}
\sum_{n\in\N}Q^E_{ns}x_{nwt}\ge E_{wts},
\qquad \forall w,t,s \text{ for which expert presence is required}.
\label{eq:expert}
\end{equation}

Nurse workload is
\begin{equation}
R_n=R_n^0+\sum_{w\in\W}\sum_{t\in\T}h_t x_{nwt},
\qquad \forall n,
\label{eq:hours}
\end{equation}
and overtime satisfies
\begin{equation}
O_n\ge R_n-C_n,\qquad O_n\ge0,
\qquad \forall n.
\label{eq:overtime}
\end{equation}
The workload spread is represented by
\begin{equation}
R^{\max}\ge R_n,\qquad R^{\min}\le R_n,
\qquad \forall n.
\label{eq:spread}
\end{equation}

\subsection*{3.3 Robot task assignment, charging, and battery dynamics}

A robot cannot work in more than one ward or work while charging:
\begin{equation}
\sum_{w\in\W}z_{rwt}+c_{rt}\le1,
\qquad \forall r,t.
\label{eq:robotexclusive}
\end{equation}
Daily robot utilization is limited by
\begin{equation}
\sum_{t\in\T_d}\sum_{w\in\W}z_{rwt}\le U_r,
\qquad \forall r,d.
\label{eq:robotday}
\end{equation}
If a minimum charging requirement is imposed,
\begin{equation}
\sum_{t\in\T_d}c_{rt}\ge m_r^C,
\qquad \forall r,d.
\label{eq:mincharge}
\end{equation}

Task execution is linked to robot presence and capability:
\begin{align}
v_{rwtl}&\le z_{rwt}, &&\forall r,w,t,l, \nonumber\\
v_{rwtl}&\le A^R_{rl}, &&\forall r,w,t,l, \nonumber\\
\sum_{w\in\W}\sum_{l\in\Lset}v_{rwtl}&\le K_{rt}, &&\forall r,t.
\label{eq:robotasks}
\end{align}
The last inequality prevents a robot from being credited with more services than it can execute during a period. Robot-service demand is covered by
\begin{equation}
\sum_{r\in\R}v_{rwtl}+u^R_{wtl}\ge D^R_{wtl},
\qquad \forall w,t,l.
\label{eq:robotneed}
\end{equation}

Battery dynamics are defined for $t=1,\ldots,|\T|$:
\begin{equation}
B_{r,t+1}=B_{rt}-\sum_{w\in\W}e_{rw}z_{rwt}+ch_r c_{rt},
\qquad \forall r,t.
\label{eq:battery}
\end{equation}
The state is bounded for every $t\in\Tp$:
\begin{equation}
0\le B_{rt}\le B_r^{\max},
\qquad \forall r,t\in\Tp.
\label{eq:batterybound}
\end{equation}
An active robot must have at least its minimum starting battery:
\begin{equation}
B_{rt}\ge B_r^{\min}\sum_{w\in\W}z_{rwt},
\qquad \forall r,t\in\T.
\label{eq:minbattery}
\end{equation}
Together, Eqs.~\eqref{eq:battery}--\eqref{eq:minbattery} also prevent a work decision that would drive the next battery state below zero or above capacity.

{\color{black}Human supervision has both a hard headcount condition and an explicit workload condition. The headcount condition remains
\begin{equation}
\sum_{r\in\R}z_{rwt}\le \rho_w^R\sum_{n\in\N}x_{nwt},
\qquad \forall w,t,
\label{eq:supervision}
\end{equation}
while supervision workload satisfies
\begin{align}
\sum_{n\in\N}\phi_{nwt} &\ge \sum_{r\in\R} d^{sup}_{rw}z_{rwt}, &&\forall w,t,\label{eq:supwork}\\
0\le \phi_{nwt} &\le \bar\phi_n x_{nwt}, &&\forall n,w,t.\label{eq:supcap}
\end{align}
Thus, robot activity consumes a calibrated fraction of nurse capacity through $\phi_{nwt}$, and Eq.~\eqref{eq:skill} no longer counts supervising nurses as fully available for clinical coverage. The coefficient $\rho_w^R$ can be lower in high-acuity wards, while $d^{sup}_{rw}$ can be estimated from observed intervention/supervision time. This remains an aggregate nurse-equivalent approximation; a task-time model would be needed if supervision must be allocated to specific clinical activities. The present formulation also assumes that any assigned nurse is eligible to provide robot supervision. If supervision requires specific training, an additional nurse--robot supervision-eligibility parameter must be introduced before clinical use.}

\subsection*{3.4 Objective functions}

The first objective measures clinical and service inadequacy:
\begin{align}
Z_1={}&\sum_{w,t,s}\beta_w\rho_t\sigma_s h^S_{wts}\nonumber\\
&+\eta_R\sum_{w,t,l}\beta_w\rho_t\tau_lu^R_{wtl}.
\label{eq:z1}
\end{align}
Ward, period, skill, and service priorities are reflected by $\beta_w$, $\rho_t$, $\sigma_s$, and $\tau_l$. In practice, critical-care skills may receive much larger shortage weights than general support activities.

The second objective measures operational burden and fairness:
\begin{align}
Z_2={}&\lambda_B(R^{\max}-R^{\min})
+\lambda_O\sum_n O_n
+\lambda_N\sum_{n,w,t\in\TN}x_{nwt}\nonumber\\
&+\lambda_P\sum_{n,w,t}q_{nwt}x_{nwt}
+\lambda_E\sum_{r,w,t}e_{rw}z_{rwt}
+\lambda_C\sum_{r,t}c_{rt}.
\label{eq:z2}
\end{align}
The model is therefore bi-objective:
\begin{equation}
\min\;(Z_1,Z_2).
\label{eq:biobj}
\end{equation}
The two objectives are deliberately kept separate because service adequacy and operating burden represent different managerial concerns.

\section*{4. SOLUTION METHODOLOGY}
The formulation is solved in two complementary ways. The first examines the trade-off between the two objectives through normalized weighted-sum and epsilon-constraint procedures, including lexicographic refinement for the reported efficient solutions. The second embeds the scheduling model in a receding-horizon framework so that updated workload, availability, demand, and robot battery states can influence subsequent decisions.
\subsection*{4.1 Multi-objective solution strategy}

Two procedures are used to examine the trade-off between $Z_1$ and $Z_2$. For a normalized weighted sum, let $Z_i^{\min}$ and $Z_i^{\max}$ be finite scaling values obtained from the individual objective anchors. Define
\begin{equation}
\widehat Z_i=
\frac{Z_i-Z_i^{\min}}
{\max\{Z_i^{\max}-Z_i^{\min},\delta\}},
\qquad \delta>0,
\label{eq:normalization}
\end{equation}
and solve
\begin{equation}
\min\;\theta\widehat Z_1+(1-\theta)\widehat Z_2,
\qquad 0\le\theta\le1.
\label{eq:weighted}
\end{equation}
The positive constant $\delta$ prevents division by zero when an anchor range vanishes and keeps the normalization well defined across instances.

The principal reporting method is the epsilon-constraint approach (Mavrotas, 2009). A shortage-priority lexicographic solution first obtains
\[
Z_1^*=\min Z_1,
\]
and then solves
\begin{equation}
\min Z_2
\qquad\text{s.t.}\qquad
Z_1\le Z_1^*+\eps.
\label{eq:eps}
\end{equation}
For the replicated benchmark, an operational reference is also useful. A pure $Z_2$ solve first returns an operationally optimal nurse--robot schedule. Because shortage and coverage-mode variables do not appear in $Z_2$, their raw values need not be unique. To obtain a reproducible shortage measure for that schedule, the assignment and robot decisions are fixed and $Z_1$ is minimized over the remaining coverage-mode and shortage variables. The resulting value is denoted $Z_1^{op}$. The benchmark then minimizes $Z_2$ subject to $Z_1\le fZ_1^{op}$. The imposed budget, rather than the raw shortage slack returned by a $Z_2$ solve, is used to state the guaranteed shortage reduction.

For Pareto analysis, each epsilon-constrained solution is additionally refined lexicographically to remove weakly efficient points caused by ties in $Z_2$. For a fixed budget $\varepsilon$, Stage~1 computes
\begin{equation}
Z_2^{*}(\varepsilon)=\min\{Z_2: Z_1\le\varepsilon\},
\label{eq:epsstage1}
\end{equation}
and Stage~2 solves
\begin{equation}
\min Z_1
\quad\text{s.t.}\quad
Z_1\le\varepsilon,\qquad
Z_2\le Z_2^{*}(\varepsilon)+\tau,
\label{eq:epsrefine}
\end{equation}
where $\tau$ is a small numerical tolerance. The two extreme points are treated analogously: the safety endpoint minimizes $Z_1$ and then $Z_2$, whereas the operational endpoint minimizes $Z_2$ and then $Z_1$. A final pairwise dominance check removes duplicates or dominated points before a frontier is reported. This procedure yields a numerically efficient set up to the stated MILP tolerance rather than relying on arbitrary auxiliary-variable values.

\subsection*{4.2 Receding-horizon predictive control}

The static model becomes a deterministic receding-horizon scheduling model when the implemented state is carried from one optimization window to the next. At control iteration $p$, let $\mathcal{T}_p$ denote the ordered set of periods in the prediction window and let $\mathcal{I}_p\subseteq\mathcal{T}_p$ denote the periods that are actually implemented before the next re-optimization (one day in the reported experiment). Let $t_p^{0}$ be the first period in the window, $t_p^{F}$ the last period in the prediction window, and $t_p^{I}$ the last implemented period. The state entering iteration $p$ is
\begin{equation}
X_p=\left(B^{0}_{rp},H_{np},N_{np},Y_{np},A_{nt},D_{wts}\right)_{t\in\mathcal{T}_p},
\label{eq:state}
\end{equation}
where $B^{0}_{rp}$ is the battery state at the beginning of the window, $H_{np}$ is cumulative implemented nurse work, $N_{np}$ is cumulative implemented night duty, $Y_{np}$ indicates whether nurse $n$ worked the immediately preceding implemented night shift, and $A_{nt}$ and $D_{wts}$ are observed or forecast exogenous inputs over the prediction window. {\color{black}Here ``prediction'' means that future demand and availability values are supplied over the finite look-ahead window; no statistical forecasting model is estimated in this paper.}

The within-window model uses the same decision variables and constraints as the static MILP, with the initial battery fixed by
\begin{equation}
B_{r,t_p^{0}}=B^{0}_{rp},\qquad \forall r.
\label{eq:mpcinitbattery}
\end{equation}
Only decisions in $\mathcal{I}_p$ are committed. Cumulative work and night-duty states are then propagated as
\begin{align}
H_{n,p+1}&=H_{np}+\sum_{t\in\mathcal{I}_p}h_t\sum_{w\in\W}x_{nwt},\label{eq:workstate}\\
N_{n,p+1}&=N_{np}+\sum_{t\in\mathcal{I}_p}I_t^N\sum_{w\in\W}x_{nwt},\label{eq:nightstate}
\end{align}
where $I_t^N=1$ for a night period and 0 otherwise. The battery state passed to the next iteration is the state immediately after the last implemented period,
\begin{equation}
B^{0}_{r,p+1}=B_{r,t_p^{I}+1},\qquad \forall r.
\label{eq:batterystateprop}
\end{equation}
The night-history indicator is updated from the last implemented shift: $Y_{n,p+1}=1$ if nurse $n$ is assigned to a night shift in period $t_p^{I}$, and 0 otherwise.

{\color{black}Cross-window rest is imposed explicitly. If $t_p^{0}$ is a morning shift that is incompatible with a night shift immediately before the current window, then
\begin{equation}
Y_{np}+\sum_{w\in\W}x_{nw t_p^{0}}\le1,\qquad \forall n.
\label{eq:crosswindow}
\end{equation}
Thus, a night assignment implemented in the preceding window cannot be followed by an incompatible morning assignment merely because the optimization horizon has moved. To mitigate artificial battery depletion at the end of the prediction window, the proposed formulation also imposes
\begin{equation}
B_{r,t_p^{F}+1}\ge B_r^{term},\qquad \forall r,
\label{eq:terminalbattery}
\end{equation}
where $B_r^{term}$ is a prespecified terminal reserve. The reserve should be calibrated from the battery required for plausible near-future service or charging access; alternatively, a soft terminal penalty can be used if a hard reserve is operationally inappropriate.}

At each iteration the current state is observed, the finite-horizon MILP is built and solved, only $\mathcal{I}_p$ is implemented, and the state is propagated before the next solve. {\color{black}The prediction horizon therefore controls how far future exogenous inputs and battery consequences can influence the current implemented decisions. A one-day window is the myopic rolling baseline; longer windows provide additional look-ahead at the cost of a larger MILP. This procedure is more accurately described as deterministic rolling-horizon re-optimization with MPC-style state propagation unless an explicit forecasting model is added.}

\section*{5. COMPUTATIONAL STUDY}
\textcolor{black}{The computational study is designed for numerical verification, comparative ablation, and controlled stress testing rather than clinical validation.} It first evaluates the reference instance under the proposed formulation, then examines synthetic instances of increasing size, multiple shortage budgets, direct baselines, parameter/weight sensitivity, and prediction-horizon effects. \textcolor{black}{Because no real hospital roster, staffing policy, or robot telemetry dataset is used, the study does not establish deployment performance or clinical benefit.}

\subsection*{5.1 Reference numerical verification}

The reference case contains eight nurses, two wards (ICU and General), three shifts, three nursing skill categories, two robots, and two robot-service types. Both robots can execute transport; only the second robot can execute the disinfection/support service. \textcolor{black}{Workload relief is credited only for an executed eligible service, and supervision consumes nurse capacity through $\phi_{nwt}$.}

\textcolor{black}{Solving this case under the proposed formulation gives $Z_1=0.78$ and $Z_2=3.164$. The entire shortage is human-coverage shortage created by the explicit supervision deduction; robot-service shortage is zero. The returned nurse plan assigns N2 and N8 to morning ICU, N4 and N6 to morning General, N1 to evening ICU, N5 to evening General, N3 to night ICU, and N7 to night General. Robot R1 performs morning ICU transport, remains idle in the evening, and charges at night; R2 charges in the morning, performs evening ICU transport, and performs the General-ward disinfection service at night. The corresponding battery paths are $(80,50,50,90)$ for R1 and $(60,100,65,35)$ for R2. All nurses work eight hours and incur no overtime.} These results provide a consistency check for the task-execution, supervision, charging, and battery constraints; they are not a clinical validation exercise.

\subsection*{5.2 Synthetic benchmark design}

The synthetic benchmark captures heterogeneous wards, skills, and robot capabilities under the proposed formulation. \textcolor{black}{Five independent seeds are reported for B20, B30, and B45; one B60 instance is retained as a larger stress case. The numerical study reports 16 instances: five seeds each for B20, B30, and B45, together with one B60 stress case.} Table~\ref{tab:sizes} summarizes the tested dimensions. Each benchmark instance uses one day with three shifts; multi-day behavior is examined separately in Section~5.6.

\begin{table*}[!t]
\centering
\caption{Synthetic benchmark dimensions.}
\label{tab:sizes}
\begin{tabular}{lrrrrr}
\toprule
Class & Nurses & Wards & Robots & Periods & Seeds \\
\midrule
B20 & 20 & 3 & 3 & 3 & 5 \\
B30 & 30 & 4 & 4 & 3 & 5 \\
B45 & 45 & 5 & 6 & 3 & 5 \\
B60 & 60 & 6 & 8 & 3 & 1 \\
\bottomrule
\end{tabular}
\end{table*}

The generator includes general care, critical care, medication competency, and senior status. Alternative coverage modes represent expert, experienced, and general qualification thresholds. Robot fleets are heterogeneous in capability, task capacity, battery state, charging rate, energy use, and support contribution. These are engineering test parameters rather than clinical recommendations.

\subsection*{5.3 Benchmark and epsilon-budget analysis}
\label{sec:revisedbenchmark}

The mixed-integer problems were solved with HiGHS (Huangfu and Hall, 2018). For each benchmark instance, an operational reference is computed and then a constrained run imposes $Z_1\le0.60Z_1^{op}$. \textcolor{black}{The reported $Z_1$ in Table~\ref{tab:agg} is the canonical achieved shortage of the returned schedule, not the imposed budget. Consequently, the difference $0.60Z_1^{op}-Z_1$ can be read as canonical budget slack.} The larger cases are computationally more difficult; therefore, status and MIP-gap information is reported rather than treating every feasible incumbent as a proven optimum.

\begin{table*}[!t]
\centering
\scriptsize
\caption{Benchmark results under the $0.60Z_1^{op}$ shortage budget (mean $\pm$ sample standard deviation where five seeds are available).}
\label{tab:agg}
\resizebox{\textwidth}{!}{%
\begin{tabular}{lccccccc}
\toprule
Class & $Z_1^{op}$ & Achieved $Z_1$ & Canonical reduction (\%) & $Z_2^{op}$ & $Z_2$ at budget & $\Delta Z_2$ (\%) & Termination \\
\midrule
B20 & $96.72\pm8.24$ & $55.29\pm4.77$ & $42.80\pm2.22$ & $47.51\pm0.38$ & $49.17\pm0.66$ & $3.50\pm1.36$ & 5 optimal \\
B30 & $116.23\pm10.58$ & $67.51\pm4.79$ & $41.83\pm1.51$ & $67.27\pm0.18$ & $70.59\pm0.65$ & $4.94\pm0.84$ & 3 optimal, 2 time-limit \\
B45 & $117.89\pm12.88$ & $68.10\pm6.08$ & $42.13\pm1.59$ & $82.50\pm3.19$ & $94.29\pm11.12$ & $14.19\pm11.69$ & 5 time-limit \\
B60 (seed 1) & 121.30 & 72.17 & 40.51 & 95.07 & 104.46 & 9.88 & time-limit \\
\bottomrule
\end{tabular}}
\end{table*}

\textcolor{black}{The 0.60 coefficient is therefore treated as a policy constraint rather than as an independently discovered improvement. In the reported runs, the canonical achieved shortage is below the imposed budget in every reported case. For B20--B45, the mean canonical reduction relative to $Z_1^{op}$ is approximately 41.8--42.8\%; the B60 seed-1 incumbent gives 40.5\%. Because several B30/B45/B60 runs terminate at the time limit, these values are reported as achieved feasible outcomes rather than universal optima. The largest time-limited MIP gap is 27.99\%, so the large-instance results are interpreted as computational stress evidence, not exact frontier points.}

\textcolor{black}{To examine sensitivity to multiple shortage-budget levels} and \textcolor{black}{to extend the epsilon analysis beyond B20}, Table~\ref{tab:epsB30} reports a B30 seed-1 sweep. The achieved shortage is canonicalized for each returned schedule. At the 0.85, 0.70, and 0.60 budgets, the canonical slack is positive, so the returned schedules lie strictly inside the corresponding policy bounds. The 0.70 run is time-limited and is therefore treated as a feasible incumbent.

\begin{table*}[!t]
\centering
\small
\caption{\textcolor{black}{Multiple epsilon budgets for B30 seed~1; achieved $Z_1$ and canonical budget slack are reported explicitly.}}
\label{tab:epsB30}
{\color{black}
\begin{tabular}{rrrrrc}
\toprule
Factor & Budget $\varepsilon$ & Achieved $Z_1$ & Slack & $Z_2$ & Solver status \\
\midrule
1.00 & 117.783 & 117.783 & 0.000 & 67.443 & optimal \\
0.85 & 100.115 & 95.940 & 4.175 & 67.443 & optimal \\
0.70 & 82.448 & 77.020 & 5.428 & 71.121 & time-limit \\
0.60 & 70.670 & 68.345 & 2.325 & 71.271 & optimal \\
\bottomrule
\end{tabular}}
\end{table*}

\begin{figure*}[t]
\centering
\begin{tikzpicture}
\begin{axis}[
width=0.76\textwidth,
height=0.43\textwidth,
xlabel={Achieved $Z_1$: weighted shortage},
ylabel={$Z_2$: operational/fairness cost},
grid=major,
]
\addplot+[mark=o] coordinates {
(68.345,71.271)
(77.020,71.121)
(95.940,67.443)
(117.783,67.443)
};
\node[anchor=south east] at (axis cs:68.345,71.271) {$0.60$};
\node[anchor=south west] at (axis cs:77.020,71.121) {$0.70^{*}$};
\node[anchor=south] at (axis cs:95.940,67.443) {$0.85$};
\node[anchor=north] at (axis cs:117.783,67.443) {$1.00$};
\end{axis}
\end{tikzpicture}
\caption{\textcolor{black}{Epsilon-budget sensitivity for B30 seed~1. Labels denote the budget factor relative to $Z_1^{op}$; the $0.70$ point (marked with an asterisk) is a time-limited feasible incumbent. The connecting segments are shown only as a visual guide and are not interpreted as a continuous Pareto frontier.}}
\label{fig:epsB30}
\end{figure*}
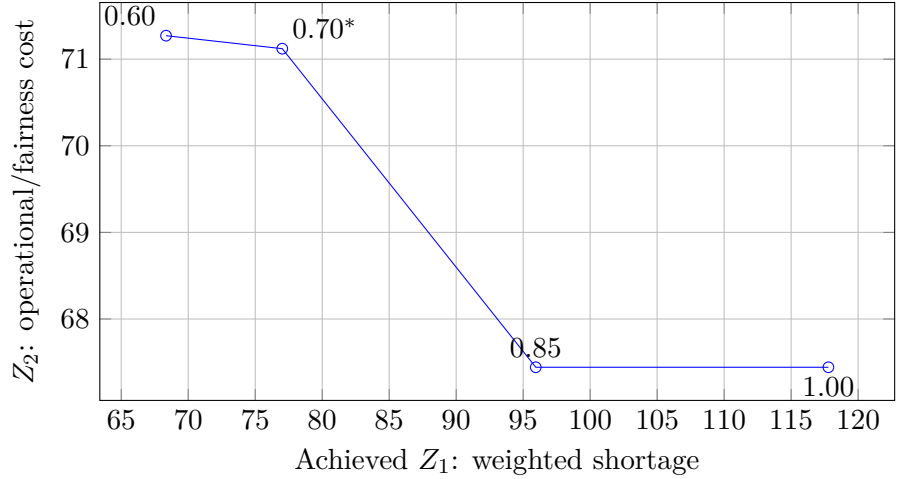

Table~\ref{tab:epsB30} and Figure~\ref{fig:epsB30} show the expected service--operation trade-off once the budget becomes tighter: reducing canonical shortage below approximately 96 increases $Z_2$. Because this is a mixed-integer problem and one intermediate point is time-limited, the table is presented as an epsilon-budget sensitivity study rather than a complete Pareto frontier.

\subsection*{5.4 Direct baseline comparison}
\label{sec:baselines}

\textcolor{black}{Three directly comparable formulations were evaluated on B20 seed~1 under the common service standard $Z_1\le60$: (i) a nurse-only model with $z=v=c=0$; (ii) a sequential model that first fixes the nurse schedule and then allocates robots; and (iii) the integrated joint formulation.} Table \ref{tab:baseline} summarizes the common-budget comparison of the nurse-only, sequential nurse-robot, and integrated joint formulations on B20 seed 1.\textcolor{black}{These controls do not reproduce every specialized algorithm in the literature, but they isolate the incremental value of robot availability and joint co-optimization under identical staffing, service, and objective definitions.}

\begin{table*}[!t]
\centering
\small
\caption{\textcolor{black}{Common-budget baseline comparison on B20 seed~1 ($Z_1\le60$).}}
\label{tab:baseline}
{\color{black}
\resizebox{\textwidth}{!}{%
\begin{tabular}{lrrrrr}
\toprule
Method & Human shortage & Robot-service shortage & Total $Z_1$ & $Z_2$ & Reported MIP gap \\
\midrule
Nurse-only & 39.250 & 20.205 & 59.455 & 50.664 & 4.99\% \\
Sequential nurse--robot & 37.825 & 20.205 & 58.030 & 51.314 & 0.00\% \\
Integrated joint & 47.263 & 6.435 & 53.698 & 50.518 & 5.00\% \\
\bottomrule
\end{tabular}}}
\end{table*}

\textcolor{black}{On this controlled instance, joint optimization reduces total weighted shortage by 9.68\% relative to nurse-only scheduling and by 7.47\% relative to the sequential approach, while keeping $Z_2$ comparable. The gain comes primarily from lower robot-service shortage; human shortage alone is not uniformly lower.} \textcolor{black}{Accordingly, the result supports the value of joint coordination on this instance but is not presented as a general superiority claim over all nurse-rostering or robot-scheduling algorithms.}

\subsection*{5.5 Parameter and objective-weight sensitivity}
\label{sec:sensitivity}

\textcolor{black}{The workload-relief and supervision coefficients are not directly observable clinical constants. In practice, $\alpha_{rls}$ should be estimated from the nurse time displaced by a completed robot service, while $d^{sup}_{rw}$ should be estimated from time-and-motion observation, robot intervention logs, or structured staff studies. To examine local sensitivity, one fixed synthetic instance was solved under a common shortage budget while scaling the relief or supervision coefficient by 0.5 and 1.5 relative to baseline.} Table \ref{tab:robotsens} reports the local sensitivity results for the robot-relief and supervision coefficients under the common shortage budget.

\begin{table*}[!t]
\centering
\small
\caption{\textcolor{black}{Local sensitivity to robot-relief and supervision coefficients under a common shortage budget.}}
\label{tab:robotsens}
{\color{black}
\begin{tabular}{lrrrr}
\toprule
Case & Achieved $Z_1$ & $Z_2$ & Reported MIP gap & Runtime (s) \\
\midrule
Relief $0.5\times$ & 82.796 & 29.561 & 0.69\% & 5.04 \\
Baseline & 82.841 & 29.690 & 1.38\% & 5.76 \\
Relief $1.5\times$ & 82.812 & 30.353 & 4.15\% & 4.09 \\
Supervision $0.5\times$ & 82.956 & 29.868 & 2.37\% & 3.80 \\
Supervision $1.5\times$ & 82.564 & 29.572 & 0.82\% & 6.67 \\
\bottomrule
\end{tabular}}
\end{table*}

\textcolor{black}{Within this local range, changes in canonical shortage are small and of the same order as the reported solver gaps, so the experiment does not support a strong monotonic sensitivity claim. Its main implication is that these parameters must be calibrated from observed task and supervision times rather than chosen solely for numerical convenience.}

\textcolor{black}{A second sensitivity test changes the relative weights in $Z_2$. Because changing the weights changes the objective scale itself, Table~\ref{tab:weightsens} reports schedule components rather than comparing the numerical $Z_2$ values across rows.}

\begin{table*}[!t]
\centering
\scriptsize
\caption{\textcolor{black}{Illustrative objective-weight sensitivity on one fixed synthetic instance.}}
\label{tab:weightsens}
{\color{black}
\begin{tabular}{lrrrrrc}
\toprule
Setting & Achieved $Z_1$ & Energy & Charges & Night shifts & Preference sum & Status \\
\midrule
Baseline & 82.841 & 63.919 & 1 & 3 & 5 & optimal \\
Fairness-heavy & 76.673 & 84.469 & 1 & 5 & 15 & optimal \\
Energy/charging-heavy & 76.080 & 62.785 & 0 & 3 & 7 & optimal \\
Preference-heavy & 83.105 & 63.919 & 0 & 3 & 12 & time-limit \\
\bottomrule
\end{tabular}}
\end{table*}

\textcolor{black}{The schedules change materially when weights are altered, confirming that the coefficients require calibration. The response is not monotone in every component because the terms interact and one setting is time-limited; no universal ranking of weight choices is inferred. In application, a defensible calibration should scale components to comparable reference ranges, elicit relative priorities from staffing and robotics stakeholders, and repeat sensitivity checks around the selected values.}

\subsection*{5.6 Prediction-horizon comparison and seven-day stress path}
\label{sec:horizon}

A seven-day deterministic stress path uses 24 nurses, two wards, and three robots, with the same controlled demand, absence, and battery disturbances used to exercise state propagation. \textcolor{black}{The one-day rolling horizon provides a myopic baseline,} \textcolor{black}{the two-day horizon provides limited look-ahead,} and \textcolor{black}{a static seven-day solve provides a full-horizon reference without feedback re-optimization. Table~\ref{tab:horizon} summarizes cumulative implemented shortage, robot energy, terminal minimum battery, and computation time.}

\begin{table*}[!t]
\centering
\small
\caption{\textcolor{black}{Prediction-horizon and static full-horizon comparison on the seven-day stress path.}}
\label{tab:horizon}
{\color{black}
\resizebox{\textwidth}{!}{%
\begin{tabular}{lrrrrrr}
\toprule
Policy & Cum. human $Z_1$ & Cum. robot $Z_1$ & Cum. total $Z_1$ & Robot energy & Terminal min. battery & Runtime (s) \\
\midrule
Rolling $H=1$ & 300.962 & 111.693 & 412.655 & 65.047 & 42.619 & 7.97 \\
Rolling $H=2$ & 301.971 & 120.992 & 422.963 & 16.543 & 62.640 & 7.92 \\
Static full horizon & 310.459 & 123.467 & 433.926 & 0.000 & 73.417 & 33.58 \\
\bottomrule
\end{tabular}}}
\end{table*}

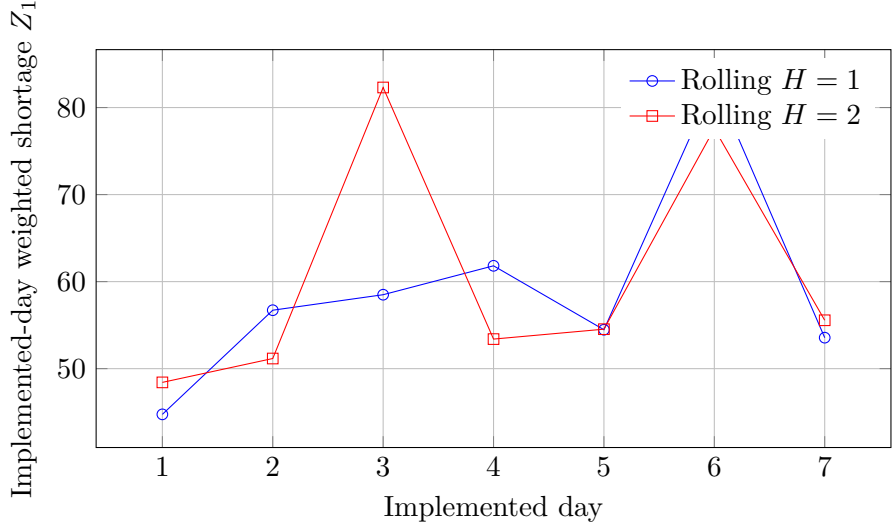
\begin{figure*}[t]
\centering
\begin{tikzpicture}
\begin{axis}[
width=0.76\textwidth,
height=0.43\textwidth,
xlabel={Implemented day},
ylabel={Implemented-day weighted shortage $Z_1$},
xtick={1,2,3,4,5,6,7},
grid=major,
legend style={draw=none,at={(0.98,0.98)},anchor=north east},
]
\addplot+[mark=o] coordinates {
(1,44.7475)(2,56.7150)(3,58.4963)(4,61.8175)(5,54.4725)(6,82.8458)(7,53.5600)
};
\addlegendentry{Rolling $H=1$}
\addplot+[mark=square] coordinates {
(1,48.4225)(2,51.1650)(3,82.3063)(4,53.3975)(5,54.5350)(6,77.5768)(7,55.5600)
};
\addlegendentry{Rolling $H=2$}
\end{axis}
\end{tikzpicture}
\caption{\textcolor{black}{Implemented-day weighted shortage trajectories for the one-day and two-day rolling horizons on the seven-day deterministic stress path. The static full-horizon policy is summarized in Table~\ref{tab:horizon} because it is solved once rather than as a sequence of daily rolling decisions.}}
\label{fig:horizontraj}
\end{figure*}

\textcolor{black}{The longer horizon does not dominate the myopic policy on shortage in this single deterministic path: $H=1$ attains the lowest cumulative $Z_1$, whereas $H=2$ preserves substantially more terminal battery and uses less robot energy. The static full-horizon solution is still more conservative in robot use and has the highest cumulative shortage. This illustrates that horizon length changes the balance between immediate service and resource conservation; it does not establish a universally optimal horizon.} \textcolor{black}{Figure~\ref{fig:horizontraj} shows the day-by-day shortage trajectories for the two rolling policies. The comparison also provides myopic and static full-horizon references under the proposed formulation.} The seven-day experiment is not intended as a probabilistic robustness assessment; such an assessment would require repeated random scenarios or an explicit uncertainty set or distribution.

\textcolor{black}{The controlled seven-day path includes a critical-demand increase, short-notice nurse absence, and a robot battery disturbance. It is used to verify state propagation and re-optimization after disturbances, not to estimate event frequencies or clinical robustness.}

{\color{black}
\subsection*{5.7 Computational complexity and scalability}
The model is a MILP and contains nurse rostering as a special case when robot variables are fixed or removed. Nurse rostering with coverage requirements and forbidden shift sequences is strongly NP-hard (den Hartog et al., 2023); therefore, the integrated problem is also NP-hard. Ignoring lower-order auxiliary variables, the principal binary-variable count is
\begin{equation}
|\N||\W||\T|+|\R||\W||\T|+|\R||\W||\T||\Lset|+|\R||\T|+|\W||\T||\Sset||\Gset|,
\label{eq:binarycount}
\end{equation}
and the supervision-load variables add $|\N||\W||\T|$ continuous variables. The dominant constraint families scale with the same products, especially robot task assignment and skill-mode coverage.}

\textcolor{black}{The one-day benchmark is therefore best viewed as preliminary resource/ward scaling rather than a proof of multi-week scalability. In the constrained runs, B20 terminates with five solver-optimal statuses under the configured tolerances; B30 has three such terminations and two time-limit incumbents; all five B45 runs and the B60 stress case reach the time limit. The largest reported time-limited MIP optimality gap is 27.99\%. This is a solver optimality gap, not the LP integrality gap. A formal LP integrality-gap analysis would require the root-relaxation values, which were not recorded in the present experiments. Therefore, a numerical LP integrality gap is not reported. Decomposition, rolling fixing, warm starts, and Benders-or logic-based separation remain future scalability directions.
}

\section*{6. DISCUSSION}

The two objectives answer different managerial questions. $Z_1$ measures inadequate clinical and support coverage, while $Z_2$ summarizes workload spread, overtime, night assignments, preferences, robot energy use, and charging. The epsilon-constraint procedure is useful when managers can state a service-shortage tolerance without defining one universal exchange rate between shortage and operational burden.

\textcolor{black}{The task-linked relief and explicit supervision formulation alters the interpretation of robot value. Robot presence alone creates no staffing credit; relief is tied to executed tasks, and supervision is charged against nurse capacity. The direct baseline comparison shows why this distinction matters: on the common-budget B20 test, the integrated model reduces total shortage mainly by reducing robot-service shortage, not by claiming that robots replace specialized nursing work.}

\textcolor{black}{The relief and supervision coefficients should be estimated from observed task completion and supervision time. The local sensitivity results should be interpreted cautiously because the observed changes are small relative to the reported solver gaps. Objective-weight sensitivity is stronger: different weight settings produce different schedules and component values. Consequently, operational deployment would require component scaling, stakeholder elicitation, and sensitivity checks rather than adopting the synthetic coefficients unchanged.}

\textcolor{black}{The seven-day comparison also clarifies the role of ``prediction.'' The model is deterministic rolling-horizon re-optimization with exogenous look-ahead values, not a learned forecasting system. Increasing the horizon from one to two days preserved more battery on the tested path but did not reduce cumulative shortage. The result is a horizon-sensitivity observation, not evidence that a particular horizon is universally superior. Likewise, the controlled disturbances do not establish stochastic robustness; that would require Monte Carlo testing, scenario-based stochastic programming, or a robust uncertainty set.}

\textcolor{black}{The battery and mobility representations remain planning-level approximations. Equation~\eqref{eq:battery} uses constant ward-period energy consumption and a constant charging increment. Robot movement is aggregated into ward-period task capacity and energy, so inter-ward routing, corridor/elevator congestion, collision avoidance, travel-time uncertainty, and charging-station occupancy are omitted. These simplifications can make a ward-level schedule optimistic when travel or charger competition consumes a material fraction of a period. Telemetry-based piecewise-linear or nonlinear battery models and explicit routing would be required for higher-fidelity deployment studies.}

\textcolor{black}{Important multi-week nurse-rostering rules are also outside the reported benchmark: maximum consecutive working days, weekly contracted-hour patterns, weekend rules, shift rotations, and longer recovery sequences are not activated. They can be added as standard inter-day constraints, but doing so changes both model size and validation requirements.} \textcolor{black}{The complexity analysis and the time-limited B45/B60 runs show why such extensions should be accompanied by larger computational studies and, potentially, decomposition or warm-start strategies.}

The evidence remains synthetic. The reference example is a numerical verification case; the larger instances are controlled engineering tests. Real deployment would require de-identified hospital data, clinically approved skill-substitution rules, institution-specific labor rules, supervision eligibility, calibrated objective weights, robot telemetry, and prospective validation with clinical and operational stakeholders.

\section*{7. CONCLUSION}

This study develops a bi-objective MILP for joint nurse and hospital-service-robot scheduling with heterogeneous nursing skills, alternative coverage modes, task-linked robot assistance, charging and battery dynamics, and explicit human-supervision workload. A deterministic receding-horizon layer carries workload, night-history, and battery states across successive optimization windows.

\textcolor{black}{The task-execution and supervision formulation yields a reference solution of $Z_1=0.78$ and $Z_2=3.164$. On a common-budget B20 seed-1 comparison, integrated joint optimization obtains total shortage $53.70$, compared with $58.03$ for sequential nurse--robot scheduling and $59.46$ for nurse-only scheduling. Multiple epsilon levels on B30 seed~1 report achieved shortage and budget slack explicitly rather than treating the imposed budget as an observed improvement.} \textcolor{black}{Sensitivity experiments show that model outputs depend on parameter and objective-weight choices, and a seven-day comparison shows different service--battery trade-offs for one-day and two-day rolling horizons.} \textcolor{black}{The larger instances also expose computational limits: several B45/B60 runs remain time-limited, so the study does not claim exact large-scale optimality or practical hospital validation.}

The results support the internal consistency and comparative value of the integrated formulation on controlled synthetic tests, while also defining the evidence still needed for deployment. Future work should use de-identified hospital data, explicit routing and charging-station constraints, telemetry-calibrated battery models, richer multi-week nurse rules, and stochastic or robust uncertainty models, together with scalable decomposition methods for larger horizons.

\section*{APPENDIX A. REFERENCE NUMERICAL DATA}
Table \ref{tab:refskills} lists the nurse-skill qualifications used in the reference example. Table \ref{tab:refreq} reports the corresponding ward- and shift-specific human coverage requirements. Table \ref{tab:refrobot} summarizes the initial battery levels, charging increments, ward-specific energy use, and service capabilities of the reference robots.
The reference case is kept separate from the synthetic benchmark generator and is used only to check the consistency of the formulation and computational results.

\begin{table*}[!t]
\centering
\caption{Nurse skills in the reference example.}
\label{tab:refskills}
\begin{tabular}{lccc}
\toprule
Nurse & General & ICU & Senior \\
\midrule
N1 & 1 & 1 & 1 \\
N2 & 1 & 1 & 1 \\
N3 & 1 & 1 & 0 \\
N4 & 1 & 0 & 1 \\
N5 & 1 & 1 & 0 \\
N6 & 1 & 0 & 0 \\
N7 & 1 & 0 & 0 \\
N8 & 1 & 0 & 0 \\
\bottomrule
\end{tabular}
\end{table*}

\begin{table*}[!t]
\centering
\caption{Human coverage requirements in the reference example.}
\label{tab:refreq}
\begin{tabular}{llccc}
\toprule
Shift & Ward & General & ICU & Senior \\
\midrule
Morning & ICU & 2.5 & 1 & 1 \\
Morning & General & 2.0 & 0 & 1 \\
Evening & ICU & 1.5 & 1 & 1 \\
Evening & General & 1.0 & 0 & 0 \\
Night & ICU & 1.0 & 1 & 0 \\
Night & General & 1.5 & 0 & 0 \\
\bottomrule
\end{tabular}
\end{table*}

\begin{table*}[!t]
\centering
\caption{Reference robot energy and capability data.}
\label{tab:refrobot}
\begin{tabular}{lrrrrcc}
\toprule
Robot & $B_0$ & Charge & ICU use & Gen. use & Transport & Disinfection \\
\midrule
R1 & 80 & 40 & 30 & 25 & 1 & 0 \\
R2 & 60 & 40 & 35 & 30 & 1 & 1 \\
\bottomrule
\end{tabular}
\end{table*}

The robot-service requirements are one transport service in ICU during the morning, one transport service in ICU during the evening, and one disinfection/support service in the General ward at night. Human supervision is set to one active robot per assigned nurse in either ward. \textcolor{black}{Under the proposed formulation, the returned robot plan is R1: ICU transport, idle, charge; R2: charge, ICU transport, General-ward disinfection. The corresponding battery trajectories are $(80,50,50,90)$ for R1 and $(60,100,65,35)$ for R2. The reference objective values are $Z_1=0.78$ and $Z_2=3.164$.}

\medskip
%\noindent\textbf{Data and code availability.} The source code and benchmark generator used for the computational study are provided as supplementary material. The larger benchmark instances are generated synthetically. No patient records or real hospital staff data are used.

\begin{simpleReferences}
\APAReference{B. Cheang, H. Li, A. Lim, and B. Rodrigues,
``Nurse rostering problems: a bibliographic survey,''
\emph{European Journal of Operational Research}, vol. 151, no. 3, pp. 447--460, 2003.
https://doi.org/10.1016/S0377-2217(03)00021-3.}

\APAReference{E. K. Burke, P. De Causmaecker, G. Vanden Berghe, and H. Van Landeghem,
``The state of the art of nurse rostering,''
\emph{Journal of Scheduling}, vol. 7, no. 6, pp. 441--499, 2004.
https://doi.org/10.1023/B:JOSH.0000046076.75950.0b.}

\APAReference{A. T. Ernst, H. Jiang, M. Krishnamoorthy, and D. Sier,
``Staff scheduling and rostering: A review of applications, methods and models,''
\emph{European Journal of Operational Research}, vol. 153, no. 1, pp. 3--27, 2004.
https://doi.org/10.1016/S0377-2217(03)00095-X.}

\APAReference{J. Van den Bergh, J. Beliën, P. De Bruecker, E. Demeulemeester, and L. De Boeck,
``Personnel scheduling: A literature review,''
\emph{European Journal of Operational Research}, vol. 226, no. 3, pp. 367--385, 2013.
https://doi.org/10.1016/j.ejor.2012.11.029.}

\APAReference{P. De Bruecker, J. Van den Bergh, J. Beliën, and E. Demeulemeester,
``Workforce planning incorporating skills: State of the art,''
\emph{European Journal of Operational Research}, vol. 243, no. 1, pp. 1--16, 2015.
https://doi.org/10.1016/j.ejor.2014.10.038.}

\APAReference{A. A. Morgan, J. Abdi, M. A. Q. Syed, G. El Kohen, P. Barlow, and M. P. Vizcaychipi,
``Robots in healthcare: A scoping review,''
\emph{Current Robotics Reports}, vol. 3, pp. 271--280, 2022.
https://doi.org/10.1007/s43154-022-00095-4.}

\APAReference{R. Huang, H. Li, R. Suomi, C. Li, and T. Peltoniemi,
``Intelligent physical robots in health care: Systematic literature review,''
\emph{Journal of Medical Internet Research}, vol. 25, e39786, 2023.
https://doi.org/10.2196/39786.}

\APAReference{J. Miseikis et al.,
``Lio: A personal robot assistant for human-robot interaction and care applications,''
\emph{IEEE Robotics and Automation Letters}, vol. 5, no. 4, pp. 5339--5346, 2020.
https://doi.org/10.1109/LRA.2020.3007462.}

\APAReference{G.-Z. Yang et al.,
``Combating COVID-19: The role of robotics in managing public health and infectious diseases,''
\emph{Science Robotics}, vol. 5, no. 40, eabb5589, 2020.
https://doi.org/10.1126/scirobotics.abb5589.}

\APAReference{M. C. Gombolay, R. A. Gutierrez, S. G. Clarke, G. F. Sturla, and J. A. Shah,
``Decision-making authority, team efficiency and human worker satisfaction in mixed human-robot teams,''
\emph{Autonomous Robots}, vol. 39, no. 3, pp. 293--312, 2015.
https://doi.org/10.1007/s10514-015-9457-9.}

\APAReference{G. Fragapane, R. de Koster, F. Sgarbossa, and J. O. Strandhagen,
``Planning and control of autonomous mobile robots for intralogistics: Literature review and research agenda,''
\emph{European Journal of Operational Research}, vol. 294, no. 2, pp. 405--426, 2021.
https://doi.org/10.1016/j.ejor.2021.01.019.}

\APAReference{E. F. Camacho and C. Bordons,
\emph{Model Predictive Control}, 2nd ed. Springer, 2007.
https://doi.org/10.1007/978-0-85729-398-5.}

\APAReference{F. Borrelli, A. Bemporad, and M. Morari,
\emph{Predictive Control for Linear and Hybrid Systems}. Cambridge University Press, 2017.
https://doi.org/10.1017/9781139061759.}

\APAReference{S. J. Qin and T. A. Badgwell,
``A survey of industrial model predictive control technology,''
\emph{Control Engineering Practice}, vol. 11, no. 7, pp. 733--764, 2003.
https://doi.org/10.1016/S0967-0661(02)00186-7.}

\APAReference{R. Zhang, F. Rossi, and M. Pavone,
``Model predictive control of autonomous mobility-on-demand systems,''
in \emph{2016 IEEE International Conference on Robotics and Automation}, pp. 1382--1389, 2016.
https://doi.org/10.1109/ICRA.2016.7487272.}

\APAReference{M. Jalaeian Farimani, R. Khalili Amirabadi, D. Nikkhouy, M. Abdolbaghi,
M. Rastegarmoghaddam, and S. Samadzadeh,
``A Systematic Review and Taxonomy of Reinforcement Learning-Model Predictive Control Integration for Linear Systems,''
\emph{arXiv preprint arXiv:2604.21030}, 2026.
https://doi.org/10.48550/arXiv.2604.21030.}

\APAReference{R. Khalili Amirabadi, M. Jalaeian Farimani, and O. Solaymani Fard,
``Experience replay in reinforcement learning: A systematic review,''
\emph{Neurocomputing}, vol. 700, Art. no. 134403, 2026.
https://doi.org/10.1016/j.neucom.2026.134403.}

\APAReference{S. J. M. den Hartog, H. Hoogeveen, and T. C. van der Zanden,
``On the complexity of Nurse Rostering problems,''
\emph{Operations Research Letters}, vol. 51, no. 5, pp. 483--487, 2023.
https://doi.org/10.1016/j.orl.2023.07.004.}

\APAReference{G. Mavrotas,
``Effective implementation of the epsilon-constraint method in multi-objective mathematical programming problems,''
\emph{Applied Mathematics and Computation}, vol. 213, no. 2, pp. 455--465, 2009.
https://doi.org/10.1016/j.amc.2009.03.037.}

\APAReference{Q. Huangfu and J. A. J. Hall,
``Parallelizing the dual revised simplex method,''
\emph{Mathematical Programming Computation}, vol. 10, no. 1, pp. 119--142, 2018.
https://doi.org/10.1007/s12532-017-0130-5.}

\end{simpleReferences}

\end{document}